\documentclass[12pt]{article}
\usepackage[onehalfspacing]{setspace}
\usepackage{amsthm}
\usepackage[hidelinks]{hyperref}
\usepackage{graphics,graphicx,amssymb,mathtools}
\usepackage{caption}
 \usepackage{booktabs}
 \usepackage{url}
 \usepackage{orcidlink}
\usepackage{dsfont}
\usepackage{xcolor}
\usepackage{bm}
\usepackage{siunitx}
\usepackage{booktabs}
\usepackage{multirow}
\usepackage{enumitem}
 \usepackage{xr} %For quoting external document
\onecolumn % for one column layouts
\usepackage{subcaption}
 \usepackage{natbib}
\usepackage{footmisc}
\usepackage{dsfont}
\usepackage{xr}
\usepackage{dsfont}
\usepackage{amssymb}
\usepackage{algorithm}
\usepackage{algorithmic}
\usepackage{array}
\usepackage{threeparttable}
  \newcommand{\ar}[1]{{{\color{orange}  #1}}}

\newcommand{\xixi}{{\xi}}
 \newcommand{\xx}{x}
 \newcommand{\ww}{w}
 
 \newcommand{\lala}{ \lambda }
  \newcommand{\gggg}{ \gamma }
 \newcommand{\argmin}{\operatornamewithlimits{arg\,min}}
 \newcommand{\vvv}{v}
\newcommand{\XX}{X}

\newtheorem{assumption}{Assumption}
\newtheorem{remark}{Remark}

\newtheorem{corollary}{Corollary}

\newtheorem{lemma}{Lemma}
\newtheorem{theorem}{Theorem}

\newtheorem{proposition}{Proposition}

\date{} 
\begin{document}
%% Here are the title, author names and addresses
\title{\bf Determining number of factors under stability considerations}

%\author{S. M. LEE\orcidlink{0009-0008-1487-2367} \and Y. CHEN\orcidlink{0000-0002-7215-2324}} %
 \author{Sze Ming Lee \orcidlink{0009-0008-1487-2367} and Yunxiao Chen \orcidlink{0000-0002-7215-2324} ~\\~\\\footnotesize\textit{Department of Statistics, London School of Economics and Political Science}}
 \maketitle
%\affil{Department of Statistics, London School of Economics and Political Science,\\ Houghton Street, London, WC2A 2AE, U.K.
%~\email{y.chen186@lse.ac.uk}}

%\affil{Department of Statistics, University of Glasgow, Glasgow G12 8QQ, U.K. \email{mike@stats.gla.ac.uk}}
\bigskip
\begin{abstract}
This paper proposes a novel method for determining the number of factors in linear factor models under stability considerations. An instability measure is proposed based on the principal angle between the estimated loading spaces obtained by data splitting. Based on this measure, criteria for determining the number of factors are proposed and shown to be consistent. This consistency is obtained using results from random matrix theory, especially the complete delocalization of non-outlier eigenvectors. The advantage of the proposed methods over the existing ones is shown via weaker asymptotic requirements for consistency, 
simulation studies and a real data example. 
\end{abstract}

\noindent%
{\it Keywords:} Data splitting; Factor analysis; Information criterion; Principal angle; Stability.
\vfill
 %
%3 to 6 keywords, that do not appear in the title

\newpage

\section{Introduction}

Factor analysis is a widely used technique for uncovering the latent structure of multivariate data. An important problem in factor analysis is to determine the number of factors in the model, which has been studied extensively in the literature (see e.g. \citealp{Bai_Ng-2002-Econometrica}, \citealp{Onatski-2009-Econometrica}, \citealp{Ahn_Horenstein-2013-Econometrica}, \citealp{Bai_etal-2018-Aos}, \citealp{dobriban_owen-2019-JRSSB} and \citealp{ke_etal-2023-JASA}). Most of the existing methods determine the number of factors by identifying a gap in the eigenvalues of the sample covariance matrix, as the factor model structure leads to some outlier eigenvalues, where the number of such eigenvalues equals the number of factors. 

This paper proposes a new method for determining the number of factors. This method is based on the concept of loading instability, which concerns the instability of the estimated loading matrix when there are multiple copies of data. As factors receive their interpretation based on the loading matrix (see e.g. \citealp{Liu_etal-2023-Psychometrika} and \citealp{Rohe_Zeng-2023-JRSSB}), this concept is important to factor analysis, especially the reproducibility of the learned factors. 
However, given the symmetric roles that loadings and factor scores play in a factor model, 
the instability of factor scores can be defined similarly, based on which the proposed method can be adapted accordingly. 
More specifically, loading instability is defined by the principal angle of two independent loading matrix estimates. Intuitively, 
the loading instability tends to be high 
when the number of factors is over-specified, as the estimated loading space contains spurious directions that correspond to singular vectors of a noise matrix. On the other hand, due to the presence of the eigengap, the loading instability tends to be low when the number of factors is correctly specified, as implied, for example, by the Davis-Kahan theorem (see e.g. \citealp{yu2015useful} and \citealp{ORourke_etal-2018-LAA}). Making use of this property, we introduce several statistical criteria for determining the number of factors and prove that they are consistent under an asymptotic regime that is not covered by those adopted in many existing methods, including \citealp{Bai_Ng-2002-Econometrica} and \citealp{Bai_etal-2018-Aos}.
The consistency is obtained using results from random matrix theory, especially the complete delocalization of non-outlier eigenvectors \citep{Bloemendal_etal-2016-PTRF}. The superiority of the proposed criteria in selecting the correct number of factors, compared to existing selection criteria, is demonstrated through simulations. Additionally, the ability of the proposed criteria to preserve loading stability is illustrated through a real data example.
 %\arr{Tried proving the result using Davis-Kahan theorem. Looks like the result only holds under quite restrictive situation. Perhaps use it as motivation without actually proving it?}
%\yc{Say a bit about the comparison between our method and the existing methods in the simulation and real data experiments.} 
%\yc{maybe also cite the paper I gave you on low-rank matrix perturbation.}

Stability is a core principle of data science \citep{yu2020inaugural}, which is important to the reproducibility of scientific discoveries. This principle has been applied to several settings of statistical learning. \cite{sun2016stabilized} defined classification instability to quantify the sampling variability of the prediction made by classification methods and proposed a stabilized nearest neighbour classifier. \cite{liu2010stability}, \cite{sun2013consistent}, \cite{yu2013stability}, and \cite{lim2016estimation} introduced stability measures for selecting tuning parameters across various statistical models, including penalized regression and high-dimensional graphical models. \cite{pfister2021stabilizing} introduced a stabilized regression algorithm designed to identify an optimal subset of predictors that generalizes across different environments.
\cite{wang2010consistent} and \cite{fang2012selection} defined stability measures for cluster analysis and used them for choosing the number of clusters. 
However, to our knowledge, the stability principle has not been applied in factor analysis. The definition of stability in factor analysis is less straightforward than that in regression due to the rotational indeterminacy of the factor model \citep{bai-2003-Econometrica,anderson1956statistical}.

%Longer version:\cite{yu2013stability} and \cite{lim2016estimation} defined estimation stability for regression models and used it to choose the tuning parameter in LASSO regression. \cite{sun2013consistent} proposed a variable selection stability measure and used it to choose the tuning parameter in penalized regression. \cite{liu2010stability} defined total instability measure for high dimensional  undirected graphs and used it for choosing the regularization parameter.

\section{Stability-based Approach}
\subsection{Factor model}
%\ar{Some random thoughts: Maybe we can describe our problem as finding the number of spikes in a spiked population model, say e.g. as in \cite{passemier2012determining} and \cite{passemier2014estimation}. Or more generally perhaps we can make a comparison with the method proposed in  \cite{ke_etal-2023-JASA}. Need to compare the assumptions required and whether our result holds for more general class of models that satisfy the assumptions in \cite{Bloemendal_etal-2016-PTRF}.} \yc{Given that the parallel analysis paper went to JRSSB, we can try that too if we can show that we can achieve consistency under a weaker condition (that I am not sure). I think JRSSB likes this type of methodology.}
We observe an $n \times p$ data matrix $X = (x_{ij})_{n \times p}$, which contains $p$ features of $n$ observations. We work with the linear factor model, satisfying  %Let $\xx_i$ be the $i$th row of $X$. 
\begin{align}\label{eq: model formulation}
    x_{ij} =  \lala_j^\top \gggg_i  + \epsilon_{ij}, 1 \leq i \leq n, 1 \leq j \leq p.
\end{align}
Here, $\gggg_i = (\gamma_{i1}, \dots,\gamma_{iK})^{\top}$ is a $K$-dimensional vector of factor scores of the $i$-th object, with $\gamma_{ik}$ being independent variables with zero mean and unit variance. $\lala_j = (\lambda_{j1},\dots, \lambda_{jK})^{\top} $ is a deterministic $K$-dimensional vector of factor loadings, and $\epsilon_{ij}$ is an independent noise term with mean $0$ and variance $\psi$. 
In matrix form, the model specified in \eqref{eq: model formulation} can be written as $X = \Gamma \Lambda^{\top} + \mathcal{E}, $
where $\Gamma = (\gamma_{ik})_{n \times K}$, $ \Lambda = (\lambda_{jk})_{p \times K}$ and $\mathcal{E}  = (\epsilon_{ij})_{n \times p}$. An intercept term can be included if entries of \( X \) are not mean zero. Since our primary goal is to determine the number of factors, and the covariance matrices are invariant to the addition of an intercept, we proceed with model \eqref{eq: model formulation} without loss of generality. The same setting has been considered in \cite{Bai_etal-2018-Aos}, under which information criteria are proposed for determining $K$. 
%\label{eq: factor model in matrix form}
%\ar{Self-reference: We need $\Sigma - I_J$ has bounded rank. So we need to stick to identical variance in error term} 
%\yc{Give a remark here to say that we can accommodate a model with intercepts.}

Let $\xx_i$ denote the $i$-th row of $X$. The population covariance matrix of $\xx_i$ is $\Sigma  = \Lambda \Lambda^{\top} + \psi I_{p}$. 
% \ar{Reference only, write in similar form on \cite{Bloemendal_etal-2016-PTRF}. }
% \begin{align*}
%     X^{\top} &=  \Lambda \Gamma^{\top} + (\Phi^{1/2})^{\top}Z^{\top} \\
%              &= (\Lambda,(\Phi^{1/2})^{\top}) \begin{pmatrix} \Gamma^{\top} \\
%              Z^{\top}
%              \end{pmatrix}.
% \end{align*}
% \begin{align*}
%     \Sigma = (\Lambda,(\Phi^{1/2})^{\top}) (\Lambda,(\Phi^{1/2})^{\top})^{\top} = \Lambda \Lambda^{\top} + (\Phi^{1/2})^{\top}\Phi^{1/2}
% \end{align*}
By eigenvalue decomposition, we can write $\Sigma = \sum_{j=1}^{p} \sigma_j \vvv_j \vvv_j^{\top},$ where $\sigma_1 \geq \dots \geq \sigma_p $ are the eigenvalues of $\Sigma$, and $\{\vvv_j\}_{j=1}^{p}$ is the set of orthonormal eigenvectors.

\subsection{Instability Measure}
We define an instability measure using the principal angle between loading matrices. Ideally, let  $U_k$  and  $V_k$  be subspaces spanned by the leading $k$ eigenvectors of the sample covariance matrices obtained from two independent and identically distributed samples. The between-sample loading instability at $k$ is defined as
\begin{align}\label{eq: population instability}
\sin \angle (U_k, V_k) = \max_{u \in U_k, u \neq 0} \min_{v \in V_k, v \neq 0} \sin \angle (u, v).
\end{align}

In practice, we only observe data from a single sample. To obtain an instability measure, we use data splitting. Let $[w]$ denote the integer part of any real number  $w$. We randomly sample the rows of $X = (\xx_1, \dotsm \xx_n)^{\top}$ without replacement to form a new permuted data matrix $(\xx_{1}^{(s)}, \dots, \xx_{n}^{(s)})^{\top}$. This data matrix is further split into two halves, $\XX^{(1)} = (\xx_{1}^{(s)}, \dots, \xx_{n_1}^{(s)})^{\top}$ and $\XX^{(2)} = (\xx_{n_1+1}^{(s)}, \dots, \xx_{n}^{(s)})^{\top}$, where  $n_1 = [n/2]$ and $n_2 = n - n_1$. 
For $l = 1, 2$, we perform eigenvalue decomposition such that $n_l^{-1} (\XX^{(l)})^{\top}(\XX^{(l)}) = \sum_{j=1}^{p} \tilde{\sigma}^{(l)}_j \tilde{\vvv}^{(l)}_j (\tilde{\vvv}_j^{(l)})^{\top}$, where $ \tilde{\sigma}^{(l)}_1 \geq \dots \geq \tilde{\sigma}^{(l)}_p $ are the eigenvalues, and  $\tilde{\vvv}_{1}^{(l)}, \dots, \tilde{\vvv}_{p}^{(l)} $ are the corresponding eigenvectors. 

Let $ \tilde{V}^{(l)}_{k} = \text{Span}\{\tilde{\vvv}^{(l)}_1, \dots, \tilde{\vvv}^{(l)}_k\} $ denote the subspace spanned by the first $k$ leading eigenvectors. The loading instability measure at $k$ is defined as
%\begin{align} \label{eq: empirical loading stability}
$\sin \angle (\tilde{V}^{(1)}_k, \tilde{V}^{(2)}_k).$
%\end{align}
When  $k = K$, this instability measure is expected to be close to zero, indicating good reproducibility of the factors. When $k > K$, the instability is expected to be close to 1, as each of $\tilde{V}^{(1)}_k$ and $\tilde{V}^{(2)}_k$ has at least one direction that corresponds to the noise matrix, resulting in two orthogonal directions. 
This phenomenon is formally stated and discussed in Section \ref{sect: Theoretical Results}.

The measure $\sin \angle (\tilde{V}^{(1)}_k, \tilde{V}^{(2)}_k)$ is computed using a single splitting of the data matrix $X$, which introduces additional randomness. To reduce this randomness, we propose to perform multiple random splittings and then take an average. Specifically, for $j = 1, \dots, J$, let  $\sin \angle (\tilde{V}^{(1,j)}_k, \tilde{V}^{(2,j)}_k)$ denote the loading instabiltiy measure computed from the $j$th split, where $J$ is the total number of splits. The averaged instability measure at $k$ is defined as $\text{INS}(k) = J^{-1} \sum_{j=1}^{J}\sin \angle (\tilde{V}^{(1,j)}_k, \tilde{V}^{(2,j)}_k).$ We use $J = 10$ for simulations and real data analysis in the rest, which seems sufficient.

\subsection{Proposed Criteria}
We propose several statistical criteria for estimating the number of factors based on the proposed instability measure. Let $\mathcal K = \{1, 2, ..., K_{\max}\}$ be a candidate set of the possible number of factors. With an appropriate decreasing deterministic sequence $\{c_k\}_{k=1}^{K_{\max}}$, whose condition is given in Theorem \ref{thm: selection for deterministic cn}, 
we can estimate $K$ consistently by
$ \argmin_{k \in \{1, \dots, K_{\max}\}}  c_k + \text{INS}(k).$
Here, $c_k \in [0,1]$ is used to prevent underestimation, 
as  
$\text{INS}(k)$ is less predictable when $k < K$.
In particular, the minimiser of 
$\text{SC1}(k) = \{{(K_{\max}-k)}/{K_{\max}}\} + \text{INS}(k)$
is a consistent estimator of $K$.  
%This demonstrates that selection criteria can be constructed using the stability measure  \eqref{eq: empirical loading stability} in a manner similar to information criteria. 

Let $\tilde{\sigma}_1 \geq \tilde{\sigma}_2 \dots \geq  \tilde{\sigma}_p$ denote the eigenvalues of 
$n^{-1} \XX^{\top}\XX. $ The following criterion estimates $K$ by combining signal strength and stability considerations:
$\text{\text{SC2}}(k)  =  {l(k)}/{ l(0)} + \text{INS}(k),$
where \(l(k) = \sum_{j=k+1}^{K_{\max}}\log(\tilde{\sigma}_j +1)\) for \(k = 0, 1, \dots, K_{\max} - 1\), and \(l(K_{\max}) = 0\). 
% The proposed estimate of the true number of factors \(K\) is 
% \[
% \hat{K}_2 = \argmin_{k \in 1, \dots, K_{\max}} \text{\text{SC2}}(k).
% \]
Here, \( l(k)/l(0) \) is analogous to the first terms of commonly used information criteria, such as the \text{AIC} and \text{BIC} proposed in \cite{Bai_etal-2018-Aos}. The denominator \( l(0) \), together with the addition of $1$ inside the logarithm to each \( l(k) \), serves to scale \( l(k)/l(0) \) within the interval \([0,1]\) to align it with the scale of the loading instability measure. This criterion aims to identify the model that balances the signal strength of the factors and their stability.
%\ar{Adjusted the proof. Running simulations again}
%\ar{Todo: Check the notation is coherent with the proof, especially SC and $\hat{K}$s once they are finalised in this section.}

%\ar{Todo: Discuss also possible criterion for a more direct comparison with existing information criterion}
Finally, we introduce a criterion related to an information criterion in  \cite{Bai_Ng-2002-Econometrica} 
%\begin{align}\label{eq: IC criterion}
   $ \text{IC}(k) = \log(p^{-1} \sum_{j = k+1}^{p} \tilde{\sigma}_j^2 ) + k g(n,p),$
%\end{align}
where $g(n,p)$ is a term depending on $n$ and $p$. 
The second term in $IC(k)$ penalizes the number of factors.
We propose a criterion in this spirit:
$\text{SC3}(k) =  { \log(1 + p^{-1} \sum_{j = k+1}^{p} \tilde{\sigma}_j^2 )}/{\log(1 + p^{-1} \sum_{j = 1}^{p} \tilde{\sigma}_j^2 )} + \text{INS}(k)$, which replaces the penalty term in the IC with the proposed instability measure. Similar to SC2, this criterion also aims to balance signal strength and stability. 
 
%\arr{Added one in both SC2 and SC3 to avoid negative values so the range of the first term lies between 0 and 1.}

% When the signal strengths do not have order $p$, \cite{Bai_etal-2018-Aos} studied the AIC and BIC criteria that can be expressed as 
% \begin{align*}
%     AIC(k) &= (p-k)\log(\bar{\sigma}_{kp}) - \sum_{j = k+1}^{p} \log(\sigma_{j}) - \frac{(p - k-1)(p-k+2)}{n} ,\\
%      BIC(k) &= (p-k)\log(\bar{\sigma}_{kp}) - \sum_{j = k+1}^{p} \log(\sigma_{j}) - \frac{(p - k-1)(p-k+2)}{2n}\log(n),
% \end{align*}
% where 
% \begin{align*}
%     \bar{\sigma}_{kp} = \frac{1}{p-k} \sum_{j = k+1}^{p}\tilde{\sigma}_{j}
% \end{align*}
% We would compare the performances of our estimators with these criteria in simulation studies. 
%Does not come up with  something too meaningfull. Maybe just compare the selection performance. 
%Self-referene. Basically by taylor expansion to calculate the difference in numerator, as well as using the assumption that all signals have the same order.

\section{Theoretical Results}\label{sect: Theoretical Results}
We provide sufficient conditions under which the selection based on \text{SC1}, \text{SC2}, and \text{SC3} is consistent. For positive sequences $\{a_n\}$ and $\{b_n\}$, we denote $a_n \lesssim b_n$ if there exists a constant $C>0$ such that $a_n \leq C b_n$ for all $n$. We denote $a_n \asymp b_n$ if $a_n \lesssim b_n$ and $b_n \lesssim a_n$. 
Further, for two sequences of random variables $A_n$ and $B_n$. We say that $A_n$ is stochastically dominated by $B_n$, if for every $\epsilon >0$ and $d>0$ there exists $N = N(\epsilon, d)$ such that $\text{pr}( A_n > n^{\epsilon} B_n) \leq n^{-d}$ for all $n \geq N$. We use the notation  \( A_n \prec B_n \) to denote that  \( A_n \) is stochastically dominated by \( B_n \). We use $O_\prec(B_n)$ to denote a term stochastically dominated by $B_n$.
The following assumptions are made. 
\begin{assumption}\label{assp: T N relation}
 $n^{1/\kappa} \leq p \leq n^\kappa$ for some positive constant $\kappa>1$. 
\end{assumption}

\begin{assumption}
\label{assp: assumption 2}
    For each integer $m \ge 1$, there exists a universal constant $\kappa_m > 0$ such that 
\begin{align}\label{eq: assp 2}
     \sup_{1 \le i \le n, 1 \le j \le p} E[|\epsilon_{ij}|^m] \le \kappa_m \text{ and } \sup_{1 \le i \le n, 1 \le k \le K} E[|\gamma_{ik}|^m] \le \kappa_m.  
\end{align}
\end{assumption}

% \begin{assump}\label{assp: FF/T converge}
% $F^{\top} F/T$ converges to a positive definite matrix as $T$ tends to infinity.
% \end{assump}
% \ar{We don't need distinct eigenvalues for the sine angle. But may need that for model fit. Check this later.}
\begin{assumption}
\label{assp: assumption 3}
 $(\sigma_K/\psi-1)(n/p)^{1/2} \to +\infty$ for $ n \gtrsim p$ and   $(\sigma_K/\psi-1)(n/p) \to +\infty$ for $p \gtrsim n$.
 % The first $K$ eigenvalues of $\Sigma$ diverges, i.e., $\sigma_K \to \infty$. \ar{Adjust after proof. Could depend on the ratio between $T$ and $N$ potentially.}
\end{assumption}

Assumption \ref{assp: T N relation} essentially requires that $\log (n)$ is comparable to $\log(p)$. Assumption \ref{assp: assumption 2} is a regularity condition on the error matrix and latent factor scores, and it can be further relaxed. For example, we can require that \eqref{eq: assp 2} holds for all $m \leq M$ for a large enough constant $M$ \citep{Bloemendal_etal-2016-PTRF}. We use the current assumption for convenience. Assumption \ref{assp: assumption 3} provides the balance between the signal strength requirement from the population covariance and the ratio requirement between the number of features $p$ and the number of samples $n$. Specifically, this is satisfied for bounded $\sigma_K$ when $n/p$ diverges, whereas a diverging $\sigma_K$ is necessary when $n \lesssim p$. When $n \asymp p$, Assumption \ref{assp: assumption 3} holds for any diverging $\sigma_K$, regardless of the convergence rate.
The following proposition characterizes the behaviour of the loading instability measure. 
% \ar{Exact version}
% \begin{proposition}\label{prop: sine angle behaviour}
%     Under Assumptions \ref{assp: T N relation} to \ref{assp: assumption 3}, we have \ar{Todo: Adjust the rate after proof.}
%     \begin{align}
%       &\sin \angle (\tilde{V}^{(1)}_{k}, \tilde{V}^{(2)}_{k} ) = 1 - O_{\prec}({(p \min\{n, p\})}^{-1/2} ) \text{ for } k > K, \label{eq: sin angle for k > K}\\
%         &\sin \angle (\tilde{V}^{(1)}_{k}, \tilde{V}^{(2)}_{k} ) = O_{\prec}(n^{-1/2} +\phi^{1/2}d_K^{-1} + d_K^{-2})   \text{ for } k = K. \label{eq: sin angle for k = K}
%     \end{align}
% \end{proposition}
%\ar{Simpler version. Avoid writing the complicated term.}
\begin{proposition}\label{prop: sine angle behaviour}
    Under Assumptions \ref{assp: T N relation} to \ref{assp: assumption 3}, we have sequences $a_n$ and $b_n$ decaying to zero, such that 
    \begin{align}
      &\sin \angle (\tilde{V}^{(1)}_{k}, \tilde{V}^{(2)}_{k} ) = 1 - O_{\prec}(a_n ) \text{ for } k > K, \label{eq: sin angle for k > K}\\
        &\sin \angle (\tilde{V}^{(1)}_{k}, \tilde{V}^{(2)}_{k} ) = O_{\prec}(b_n)   \text{ for } k = K. \label{eq: sin angle for k = K}
    \end{align}
    Additionally, if $\sigma_1, \dots, \sigma_K$ are distinct, then \eqref{eq: sin angle for k = K} also holds for $k = 1, \dots, K-1$. 
\end{proposition}
The exact expressions of $a_n$ and $b_n$ are provided in Section \ref{sect: Proof} of the Supplementary Material. The cone concentration results for outlier eigenvectors and the delocalization results for non-outlier eigenvectors from random matrix theory \citep{Bloemendal_etal-2016-PTRF} are crucial to proving Proposition \ref{prop: sine angle behaviour}. Although the Davis-Kahan theorem (see \citealp{yu2015useful} and \citealp{ORourke_etal-2018-LAA}) also provide results on principal angles, it can only prove \eqref{eq: sin angle for k = K} when $\sigma_K$  grows faster than the spectral norm of the error matrix, which is stronger than what is required in Theorem~\ref{thm: selection for deterministic cn}.
The results for \(k < K\) are provided for independent theoretical interest. They are not used for proving Theorem~\ref{thm: selection for deterministic cn}.
%Building on this proposition, we present the main theorem, which allows us to construct consistent estimators of $K$, without assuming distinct eigenvalues:
\begin{theorem}\label{thm: selection for deterministic cn}
Under Assumptions \ref{assp: T N relation} to \ref{assp: assumption 3}, for any decreasing sequence $\{c_k\}_{k=1}^{K_{\max}}$ with $1 \geq c_k \geq 0$ for $k \in \mathcal{K}$, such that for some $\delta>0$, $c_{k} - c_{k+1} >  \delta$ for all $k \in \{1, \dots, K-1\}$, and $1- \delta> c_{K} - c_{K_{\max}}$. Define 
$\tilde{K} = \argmin_{k \in \{1, \dots, K_{\max}\}}  c_k + \text{INS}(k).$
We have $\lim_{n \to \infty } \text{pr} \left( \tilde{K}  = K \right) =1.$ Consequently, \text{SC1} can consistently estimate $K$.
%\yc{make it a theorem. Add inside this theorem that \text{SC1} is consistent.}
\end{theorem}
 The corollaries below give the conditions for $\text{SC2}$ and $\text{SC3}$ to be consistent. 

\begin{corollary}\label{cor 1}
Under Assumptions \ref{assp: T N relation} to \ref{assp: assumption 3}, if additionally $p \asymp n$ and $\log \sigma_1 / \log \sigma_K \lesssim C$ for some constant $C>0$, then \text{SC2} can consistently estimate $K$
% \begin{align*}
%     pr(\hat{K} = K^*) \to 1 
% \end{align*}
as $n, p \to + \infty $. %\yc{Make it a corollary.}
\end{corollary}

\begin{corollary}\label{cor 2}
Under Assumptions \ref{assp: T N relation} to \ref{assp: assumption 3}, if additionally  $p \asymp n$ and
$\sigma_{k}^2 \asymp p$, $k = 1, \dots, K$, then \text{SC3} can consistently estimate $K$
% \begin{align*}
%     pr(\hat{K} = K^*) \to 1 
% \end{align*}
as $n, p \to + \infty $. %\yc{Make it a corollary.}
\end{corollary}
We briefly discuss the assumptions underlying the proposed criteria and compare them with existing methods. \text{SC1} is the least restrictive one, requiring only Assumptions \ref{assp: T N relation} to \ref{assp: assumption 3}. \text{SC2} and \text{SC3} require the additional condition  $p \asymp n$, which is also needed for some  other existing selection criteria, such as those in \cite{Bai_Ng-2002-Econometrica} and \cite{Bai_etal-2018-Aos}.
For \text{SC2}, it is further required that  $\log \sigma_1$  and  $\log \sigma_K$  are comparable, ensuring the gap  $\{l(k-1) - l(k)\}/l(0)$  to be bounded below by a positive constant for \( k = 1, \dots, K \). Finally, \text{SC3} imposes the strictest condition, requiring that $ \sigma_{k}^2 \asymp p $, a crucial assumption for this type of information criterion, as in \cite{Bai_Ng-2002-Econometrica}.

%\yc{Need to compare our conditions with the existing ones, and somehow shows that our asymptotic regime is less restricted. }\ar{Perhaps can demonstrate the need of $p \asymp n$ for SC2, SC3 as well as other existing selection criteria}

%\yc{What is our assumption on the eigenvalues? How large do we need the eigengap to be? }
% \ar{Perhaps we can prove a more general version first, e.g. for every reasonable decreasing sequence  $a_K$, $e.g. a_K =  -K/K_{max}$, $a_K+ \sin \angle (\tilde{V}^{(1)}_{K}, \tilde{V}^{(2)}_{K} ) $ is maximised at $K = K^*$. That way the stability measure feels stronger than just a mere penalty term and we can say the performance is best by using model fit for interpretation/theoretical result. }\yc{Agree! It looks like we should be able to show that we can achieve consistency under a weaker condition than Bai and Ng. }

\section{Numerical Experiments}\label{sect: Numerical Experiments}
\subsection{Simulation Settings}
We assess the finite sample performance of the proposed method via Monte Carlo simulations. In particular, we consider sample size $n =1,500$, where $p$ takes values from $500, 1,000, 1,500$ and $2,000$. Under each setting, 100 replications are generated. We set the true number of factors to be $K=4$, and the candidate set be $\mathcal K = \{1, 2, ..., 10\}$ for model selection. 

We simulate data from the model $X = \Gamma \Lambda^{\top} + D_{\epsilon} Q_{\epsilon}\mathcal{E}$, where the entries of $\Gamma$ are independently drawn from the uniform distribution $U[-0.5, 0.5]$. The factor loadings $\Lambda$ are generated as $\Lambda = QD$, with $D$ a diagonal matrix whose diagonal elements are $\mu_1, \dots, \mu_K$. The matrix $Q$ is orthonormal, obtained via QR decomposition of a random matrix $Z \in \mathbb{R}^{p \times K}$, where each entry of $Z$ is independently sampled from a standard normal distribution $N(0,1)$.

For the error term, we consider two scenarios to generate homogeneous and heterogeneous errors, respectively. In the first scenario (S1), $\mathcal{E}$ is drawn from $N(0,1)$, with $D_{\epsilon} = Q_{\epsilon} = I_{n}$, ensuring homogeneous errors. In the second scenario (S2), $\mathcal{E}$ is drawn from a Student’s $t$-distribution with 10 degrees of freedom. To introduce heteroskedasticity and test the robustness of the criteria under violation of the homoscedasticity assumption, $Q_{\epsilon}$ is an $n \times n$ orthonormal matrix generated similarly to $Q$, and $D_{\epsilon} = \text{diag} \left\{1/n, 2/n, \dots, 1\right\}$  is a diagonal matrix.
We also consider three sets of $\mu_j$ values, the diagonal elements of $D$,  corresponding to different signal strength settings: (i) $\{6p^{1/2}, 5p^{1/2}, 4p^{1/2}, 3p^{1/2}\}$ for strong signals, (ii) $\{6p^{1/6}, 5p^{1/6}, 3p^{1/6}, 3p^{1/6}\}$ for weak signals, and (iii) $\{3 p^{1/3}, 3 p^{1/3}, 3 p^{1/6}, 3 p^{1/6}\}$ for signals of varying strengths. Note that under the last two settings, the top eigenvalues of $\Lambda$ are not distinct. 

\subsection{Results}
\begin{figure}[]
    \centering
    \includegraphics[width=0.7\linewidth]{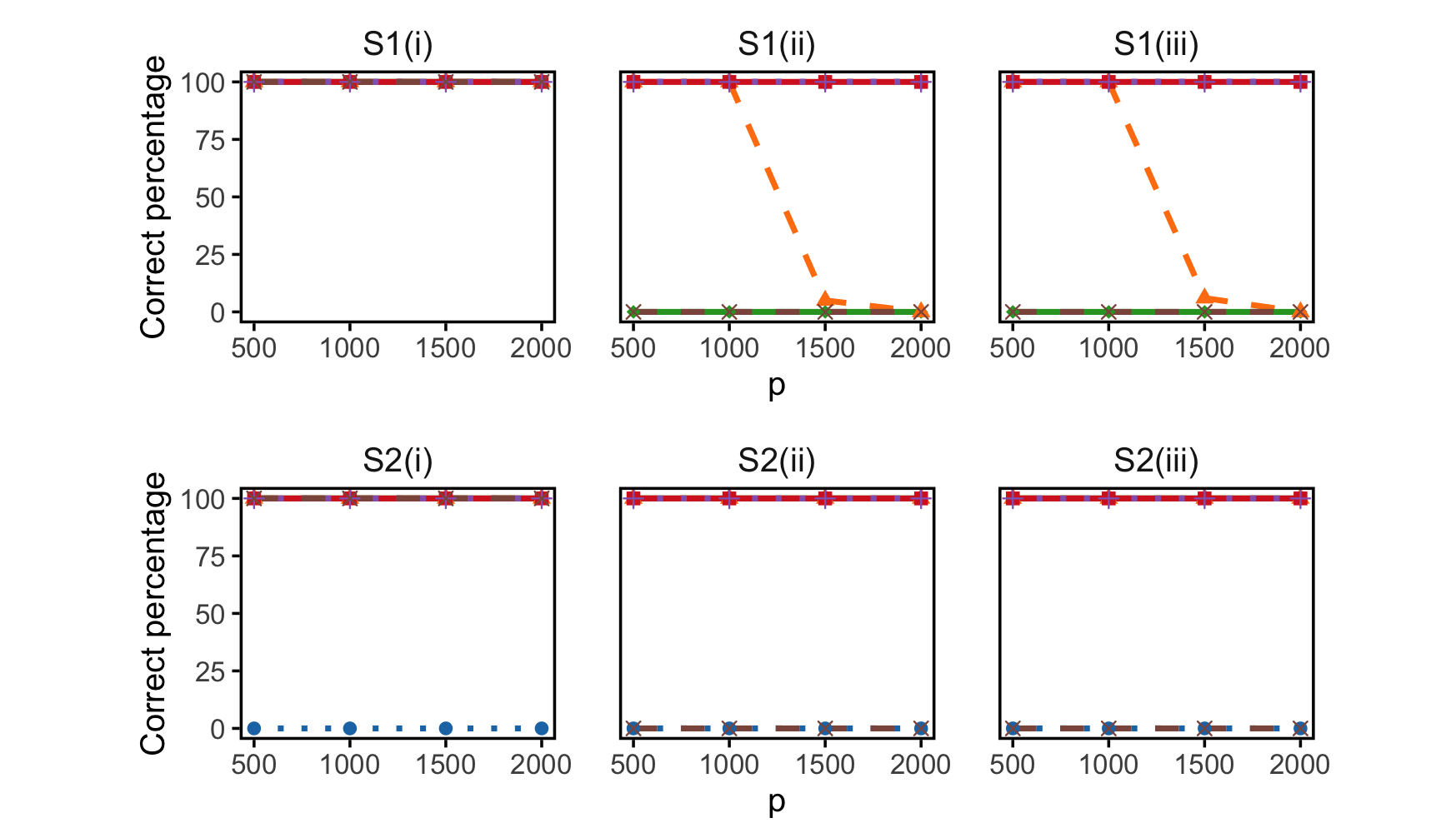}
   %  {.}
     % The first row (S1) and the second row (S2) correspond to Simulation 1 and Simulation 2, respectively. Each column (i, ii, iii) represents different signal strengths. The sample size $n$ is fixed at 1500 across all settings.
 %     \figuresize{.2}
 % \figurebox{20pc}{26pc}{}[comparemethods]
 \captionsetup{font=footnotesize} 
\caption{Correct selection percentages versus the number of features $p$ across different scenarios (S1 and S2) and signal strengths (i,ii and iii). AIC (blue dotted line with circles), BIC (orange dashed line with triangles), IC (green solid line with diamonds), SC1 (red solid line with squares), SC2 (purple dotted line with pluses), and SC3 (brown dashed line with crosses)}
    \label{fig:compare_method}
\end{figure}

We compare the stability-based criteria \text{SC1}, \text{SC2}, and \text{SC3} proposed in this work with \cite{Bai_Ng-2002-Econometrica}'s information criterion (IC) with $g(n,p) = \{(n+p)/(np)\}\log(np/(n+p))$,  and \cite{Bai_etal-2018-Aos}'s 
AIC and BIC. Figure \ref{fig:compare_method} shows the percentage of correct selections of the true number of factors $K$ by these criteria.
All methods perform well under S1(i), where the errors are homogeneous and the signal is strong. However, as expected, \text{SC3} and IC struggle when the signal does not follow the $p^{1/2}$ order, as seen in the second and third columns of the figure. AIC also underperforms when faced with heterogeneous errors, as demonstrated in all S2 scenarios. While BIC shows robustness in S2, it performs poorly under S1(ii) and S1(iii).
Overall, \text{SC1} and \text{SC2} consistently select the correct number of factors across all settings, demonstrating superior performance in both homogeneous and heterogeneous error conditions.

\begin{figure}[]
\centering
    \includegraphics[width=0.7\linewidth]{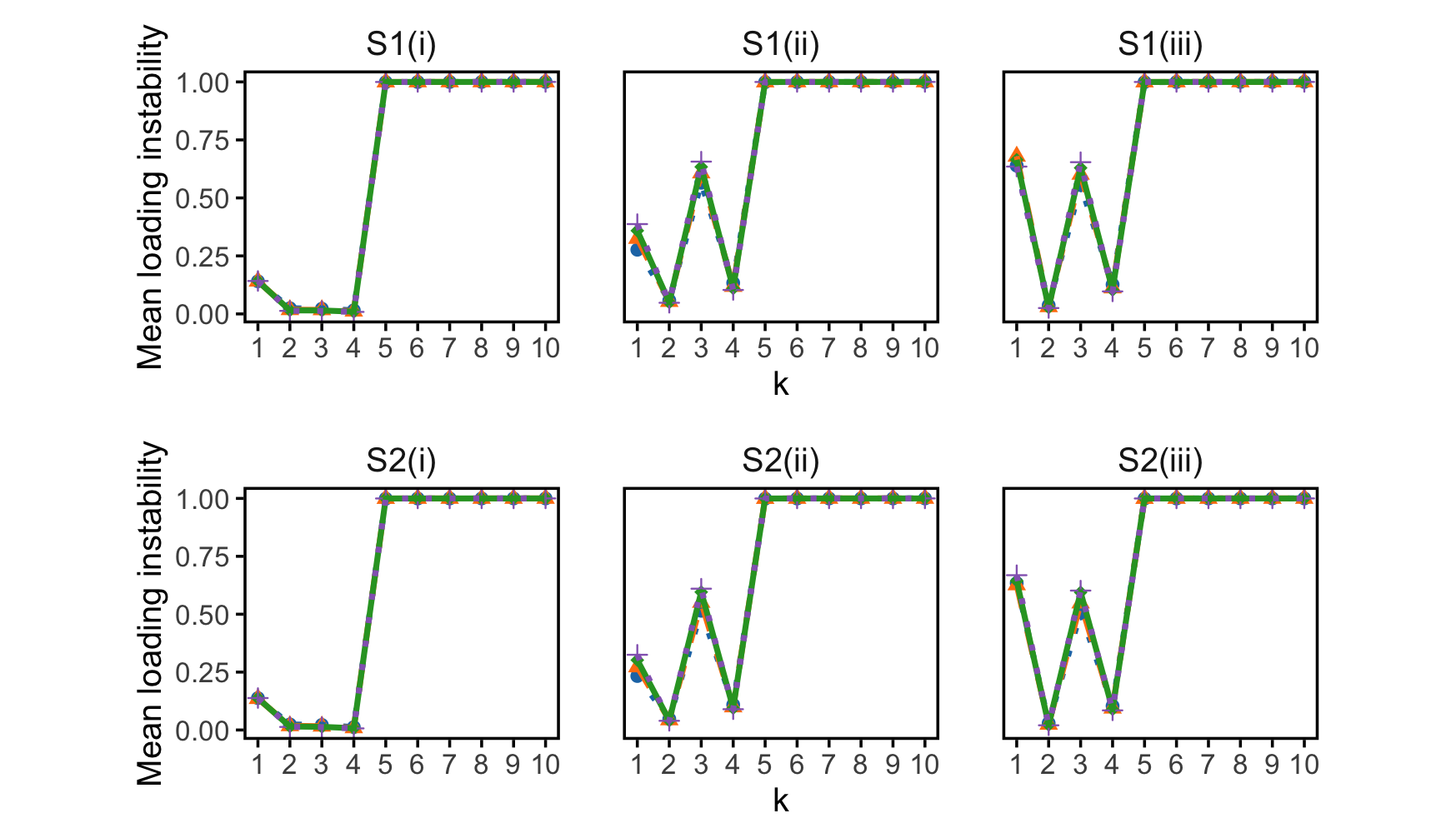}
% \figuresize{.2}
% \figurebox{20pc}{25pc}{}[stability]
\captionsetup{font=footnotesize} 
     \caption{Loading instability versus $k$ across different scenarios (S1 and S2) and signal strengths (i,ii and iii). $p=500$ (blue dotted line with circles), $p=1000$ (orange dashed line with triangles), $p=1500$ (green solid line with diamonds) and $p=2000$ (purple dotted line with pluses).} 
     %{}
     %The first row (S1) and the second row (S2) correspond to Simulation 1 and Simulation 2, respectively. Each column (i, ii, iii) represents different signal strengths. The sample size $n$ is fixed at 1500 across all settings.}
    \label{fig: stability}
\end{figure}

Figure \ref{fig: stability} illustrates the behaviour of the mean of the proposed instability measure across all replications for $k \in \mathcal{K}$. The instability measure is near 1 for $k > 4$ and close to 0 for $k = 4$, providing numerical support for Proposition \ref{prop: sine angle behaviour}. For $k = 1, 2, 3$, the instability measure is also close to 0 in the first column, where signals have distinct values. In the second and third columns, some measures deviate from 0 due to the presence of equal signal strengths.
%------------------------------------------
%------------------------------------------
%------------------------------------------
%------------------------------------------

%------------------------------------------
%\ar{Todo: Adjust stability graph and caption after changing definition }
%------------------------------------------
%------------------------------------------
%------------------------------------------
\section{Data Analysis}\label{sect: Data Analysis}
We consider a dataset concerning the p53 tumour suppressor protein, which plays a crucial role  
in cancer treatment research. This dataset includes 2D electrostatic, surface-based features, and 3D distance-based features, extracted using a method by \cite{Danziger_etal-2006-IEEE}, 
for a large collection of p53 mutations. It contains $5,208$ features for each of the $31,158$ mutations.  It is important to understand the dependence between  the features 
(see  \citealp{Lopes_etal-2019-Biometrika}).

To evaluate the performance of the proposed criteria, we focus on the first $p = 1,000$ features and sample $n = 3,000$ rows without replacement from the data matrix. 
The row sampling is performed 100 times, resulting in 100 datasets that may be regarded as independent copies of each other.  For each dataset, the features are standardized to have mean zero and variance one. The methods discussed in Section \ref{sect: Numerical Experiments} are then applied to these datasets. Table \ref{tab:example} summarises the results, including the mode of the estimated number of factors, the selection percentage of the mode, and the mean between-sample loading instability. Specifically, the mean between-sample loading instability, which measures the reproducibility of the estimated factor structure,
is calculated as follows. For every two datasets, we calculate the principal angle between the estimated loading spaces, which is well-defined even when they are of different dimensions. We then average the principal angle across all the 4,950 pairs of datasets. %\yc{Need to discuss here.}

The results show that the IC, AIC, and BIC criteria consistently estimate the number of factors to be $10$, the upper bound of our candidate set. They continue to choose the maximum number in the candidate set when we increase it to $20, 30, 40$ and $50$. The between-sample loading instability is 0.92, meaning the resulting ten-factor models have some unstable factors. 
On the other hand, all the proposed stability-based criteria tend to select two factors, with a corresponding mean between-sample loading instability being 0.10, indicating a higher level of stability. While the true number of factors is unknown, this result shows that the proposed method is more conservative than the existing criteria, which ensures a more reproducible factor structure.

\begin{table}[]
\centering
\caption{Performance of selection criteria on the p53 protein dataset}{
\begin{tabular}{lccc}
Criterion & Mode & Selection percentage$(\%)$ & Mean between-sample loading instability  \\ 
\text{SC1}   & 2  & 97 & 0.10 \\
\text{SC2}   & 2  & 98 & 0.10 \\
\text{SC3}   & 2  & 100 & 0.10 \\
IC    & 10 & 100 & 0.92 \\
AIC   & 10 & 100 & 0.92 \\
BIC   & 10 & 100 & 0.92 
\end{tabular}
}\label{tab:example}
\captionsetup{font=footnotesize} 
\caption*{Mode: mode of the estimated number of factors for each criterion. Selection percentage$(\%)$: Percentages of instances selecting the mode. Mean between-sample loading instability: mean of the principal angles between the estimated loading spaces of all 4,950 pairs of datasets.
}
\end{table}
%\ar{Todo: Adjust description to last column}}
\section{Discussion}
% In this paper, we proposed a loading stability measure for determining the number of factors in factor analysis. Using results from random matrix theory, we showed that the proposed stability-based criteria are consistent. Compared with many existing methods, the proposed one focuses directly on the stability of the estimated loading matrix and, thus, may give more reproducible results. 

Although our method is proposed under a linear factor model, the principal angle between loading spaces can be computed using two loading matrix estimates from data splitting, regardless of the specific factor model and estimator. 
Therefore, the same statistical criteria can be applied to determine the number of factors for other factor models, such as the generalized latent factor model \citep{chen_Li-2022-Biometrika} that can be used to analyze binary, categorical, and count data.
Specifically, we believe that a similar consistency result holds when the instability measure is constructed based on the constrained joint maximum likelihood estimator \citep{Chen_Li_Zhang-2020-JASA,chen_Li-2022-Biometrika} under the generalized latent factor model. However, establishing such results is nontrivial. The key to our consistency result is establishing the delocalization of the spurious directions of the estimated loading space. As the random matrix theory we currently use cannot be directly applied here, new technical tools are needed to establish the delocalization and further the consistency of the stability-based criteria. We leave it for future investigation. 

\section*{Supplementary material}
The Supplementary Material provides additional simulation and real data results and technical proofs of Proposition~\ref{prop: sine angle behaviour}, Theorem~\ref{thm: selection for deterministic cn}, and Corollaries 1 and~2. An implementation of the proposed selection criteria
for R is available at \href{https://github.com/Arthurlee51/DNFSC}{https://github.com/Arthurlee51/DNFSC}. The real data in Section \ref{sect: Data Analysis} is available from   \href{https://archive.ics.uci.edu/dataset/188/p53+mutants}{\url{https://archive.ics.uci.edu/dataset/188/p53+mutants}}. 
% \clearpage

\appendix
\renewcommand{\theequation}{S.\arabic{equation}}
\renewcommand{\thefigure}{A\arabic{figure}} % Add this line to change the figure numbering style
\renewcommand{\thetable}{A\arabic{table}}
\setcounter{equation}{0}
\setcounter{table}{0}
\setcounter{figure}{0}

\newpage  % Start a new page
%\vspace*{2cm}  % Add vertical space
\begin{center}
    \Large \textbf{Supplementary Materials for: Determining number of factors under stability considerations} \\
    % \large Your Title Here \\
    % \large Your Name \\
    % \large \today
\end{center}
\vspace{1cm}  % Add vertical space

% \title{Supplementary Materials for: Determining Number of Factors under Stability Considerations}
% \author{Sze Ming Lee\footnote{Department of Statistics, London School of Economics and Political Science, London, UK} \and Yunxiao Chen\footnotemark[1]}
% \date{}

% \maketitle
% \appendix

\section{Additional Simulation and Real Data Results}
\begin{figure}[t]
    \centering
    \includegraphics[width=0.7\linewidth]{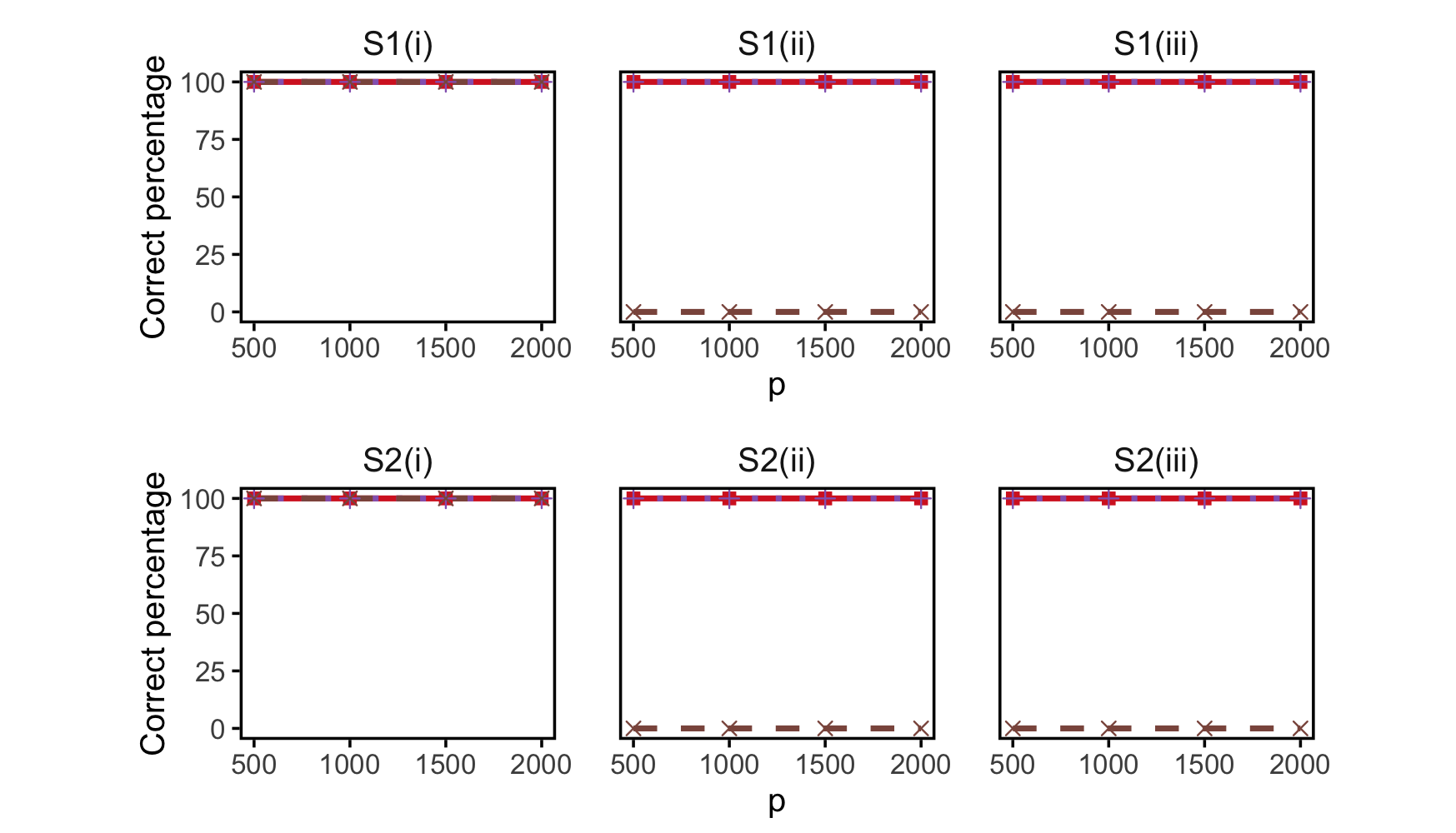}
   %  {.}
     % The first row (S1) and the second row (S2) correspond to Simulation 1 and Simulation 2, respectively. Each column (i, ii, iii) represents different signal strengths. The sample size $n$ is fixed at 1500 across all settings.
 %     \figuresize{.2}
 % \figurebox{20pc}{26pc}{}[singleselect]
 \captionsetup{font=footnotesize}
\caption{Correct selection percentages versus the number of features $p$ across different scenarios (S1 and S2) and signal strengths (i,ii and iii). All criteria are evaluated using $J=1$ for the calculation of $\text{INS}(k)$.  SC1 (red solid line with squares), SC2 (purple dotted line with pluses), and SC3 (brown dashed line with crosses)}
    \label{fig:single select}
\end{figure}

We present additional simulation results for the proposed criteria \textsc{SC1}, \textsc{SC2} and \textsc{SC3}, where the total number of splits $J$ is set to 1. Under this setup, only one random split is performed to compute the instability measure $\text{INS}(k)$. Figure \ref{fig:single select} shows the percentage of correct selections of the true number of factors $K$ by these criteria. The result demonstrates that the correct selection percentages of the proposed criteria with a single split are nearly identical to those obtained at $J=10$, as shown in figure \ref{fig:compare_method}. 

This observation is further supported by  Table \ref{tab:example single}, which presents the modes of the selected number of factors and the corresponding selection percentages for each criteria in the analysis of the p53 protein dataset using a single split.  As shown in the table, all criteria with a single split consistently select $2$ as the mode, which matches the result in Table \ref{tab:example} at $J = 10$. 

These findings suggest that the randomness introduced by splitting does not have severe impact on the performance of the proposed selection criteria. Therefore, using $J=10$ or any other reasonable number of splits is adequate in practice.

\section{Proof}\label{sect: Proof}
With slight abuse of notation, we assume $n_1 = n_2 = n$ in this section for convenience. To simplify notation, we assume $\psi = 1$ without loss of generality. Recall that the population covariance matrix is 
\begin{align*}
\Sigma = \Lambda^{\top} \Lambda + I_p = \sum_{j=1}^{p} \sigma_j \vvv_j \vvv_j^{\top}.
\end{align*}
We express the eigenvalues $\sigma_j$ of $\Sigma$ as
\begin{align*}
\sigma_j = 1 + \phi_n^{1/2}d_j, j = 1, \dots, p,
\end{align*}
where $\phi_n = p/n$ is the dimensional ratio. For notational convenience, we suppress the dependence on $n$ and write $\phi$. It is easy to see that $\sigma_j = 1$ and $d_j = 0$ for $j > K$.   
For any vector $\ww \in \mathbb{R}^{T}$, we use $w_j = \langle \vvv_j, \ww \rangle, j = 1, \dots, p$ to denote the components of $\ww$ in the eigenbasis of $\Sigma$. 
\subsection{Proof of Proposition \ref{prop: sine angle behaviour}}
The following lemma establishes a delocalisation bound for the $(K+1)$-th to $K_{\max}$-th eigenvectors of the sample covariance matrix of $X^{(l_1)}$ with respect to all eigenvectors of the sample covariance matrix of $X^{(l_2)}$ for the candidate set $\mathcal{K}$, where $l_1 \neq l_2$.

\begin{table}[t]
\centering
\caption{Performance of the Proposed Criteria ($J=1$) on the p53 Protein Dataset}
\begin{tabular}{lcc}
Criterion & Mode & Selection percentage$(\%)$ \\
\textsc{SC1}   & 2  & 95 \\
\textsc{SC2}   & 2  & 95  \\
\textsc{SC3}   & 2  & 98 
\end{tabular}
\captionsetup{font=footnotesize}
\caption*{Mode: mode of the estimated number of factors for each criterion. Selection percentage$(\%)$: Percentages of instances selecting the mode. }
\label{tab:example single}
\end{table}

\begin{lemma}\label{lm: orthogonality}
Under Assumption \ref{assp: T N relation} to \ref{assp: assumption 3}, for any $k \in \{K+1, \dots, K_{\max}\}$ and $s \in \mathcal{K}$,  we have 
\begin{align*}
 \langle\tilde{v}^{(1)}_k,\tilde{v}^{(2)}_s \rangle^2 \prec {(p \min\{n, p\})}^{-1/2} \text{ and } \langle\tilde{v}^{(2)}_k,\tilde{v}^{(1)}_s \rangle^2 \prec {(p \min\{n, p\})}^{-1/2}.
\end{align*}
\end{lemma}
Proof: We prove the first statement; the proof for the second is identical. For any deterministic unit vector $\ww$, by Theorem 2.17 of \cite{Bloemendal_etal-2016-PTRF}, we have 
\begin{align}\label{eq: thm 2.17 of bloemendal etal}
\langle \tilde{v}^{(1)}_k,\ww \rangle^2  \prec  \frac{\|\ww\|^2}{p} + \sum_{j=1}^{p} \frac{\sigma_j w_j^2}{p { (d_j - 1)^2 + \kappa_j }},
\end{align}
where $\kappa_j = \min\{n , p\}^{-2/3} \min\{ j , \min\{n , p\} + 1 - j\}^{2/3}.$ Since $d_j=0$ and $\sigma_j=1$ for $j > K$, we have
\begin{align}\label{eq: sum from K+1 to T}
    \sum_{j=K+1}^{p} \frac{\sigma_j w_j^2}{p\{(d_j-1)^2 + \kappa_j  \}} \leq  \sum_{j=K+1}^{p}\frac{ w_j^2}{p}  \leq \frac{1}{p}. 
\end{align}
 On the other hand, for $1 \leq j \leq K$, by Assumption \ref{assp: assumption 3}, we have $d_j \to +\infty$.
 Therefore, we have 
\begin{align}\label{eq: bound for i less than or equal to K}
    \frac{\sigma_j w_j^2}{p\{(d_j-1)^2 + \kappa_j  \}} &= \frac{(1 + \phi^{1/2}d_j) w_j^2}{p\{(d_j-1)^2 + \kappa_j  \}} \lesssim \frac{1}{ (np)^{-1/2}d_j }.
\end{align}
Combining equations \eqref{eq: thm 2.17 of bloemendal etal}, \eqref{eq: sum from K+1 to T}, and \eqref{eq: bound for i less than or equal to K}, we have
\begin{align}\label{eq: bound for deterministic w}
     \langle\tilde{v}^{(1)}_k,\ww \rangle^2  \prec   {(p \min\{n, p\})}^{-1/2} 
\end{align}
for any deterministic unit vector $\ww$. Note that $\tilde{v}^{(1)}_k$ and $\tilde{v}^{(2)}_s$ are independent. Therefore, by conditioning on $\tilde{v}^{(2)}_s$, we can apply \eqref{eq: bound for deterministic w} and hence the lemma is proved.  %\ar{There are rooms to relax the condition to get a convergent rate in this part. However the theorem for outlier holds with $|d_i| \lesssim (1 + N^{-1/3}) $ somehow. Needs to check again. }
%---------------------------------------------------------------------------------------------------------------------------------------------------------------------------------
%---------------------------------------------------------------------------------------------------------------------------------------------------------------------------------
%---------------------------------------------------------------------------------------------------------------------------------------------------------------------------------

We now turn to the proof of \eqref{eq: sin angle for k > K} in Proposition \ref{prop: sine angle behaviour}. It is sufficient to show that for $k > K$,
\begin{align}\label{eq: min max result for lemma}
    \min_{\xixi^{(1)} \in \tilde{V}^{(1)}_{k} ; \|\xixi^{(1)}\|=1} \max_{\xixi^{(2)} \in \tilde{V}^{(2)}_{k} ; \|\xixi^{(2)}\|=1 } \langle\xixi^{(1)},\xixi^{(2)}\rangle  \prec {(p \min\{n, p\})}^{-1/2} . 
\end{align}
Note that Lemma \ref{lm: orthogonality} implies
\begin{align*}
  \max_{\xixi^{(2)} \in \tilde{V}^{(2)}_{k} ; \|\xixi^{(2)}\|=1 } \langle\tilde{\vvv}^{(1)}_{K+1},\xixi^{(2)}\rangle \prec {(p \min\{n, p\})}^{-1/2}.
\end{align*}
Since $\tilde{\vvv}^{(1)}_{K+1} \in \tilde{V}^{(1)}_{k} $ and $\|\tilde{\vvv}^{(1)}_{K+1}\|=1,$ the proof is complete by the definition of \eqref{eq: min max result for lemma}.  \\
We now proceed to prove \eqref{eq: sin angle for k = K} in Proposition \ref{prop: sine angle behaviour}. First, we introduce some definitions and supporting lemmas. Let $A \subseteq \{1, \dots, p\}$ be a subset of integers from $1$ to $p$. For $l= 1,2$, define the random spectral projection
\begin{align*}
    P^{(l)}_A = \sum_{k \in A} \tilde{v}_k^{(l)}(\tilde{v}_k^{(l)})^{\top}.
\end{align*}
For $k= 1, \dots, p$, we also define 
\begin{align*}
    \nu_k(A) = \begin{cases}
    \min_{j \notin A}|d_k - d_j| \text{ if } k \in A\\
    \min_{j \in A}|d_k - d_j| \text{ if } k \notin A
    \end{cases}.
\end{align*}
Here $\nu_k(A)$ is the distance from $d_k$ to either $\{d_j\}_{j \in A}$ or $\{d_j\}_{j \notin A}$, whichever it does not belong to. We further define the deterministic positive quadratic form 
\begin{align*}
    \langle\ww, Z_A\ww\rangle&= \sum_{j \in A}\mu(d_j)w_j^2, \text{ where }\\
    \mu(d_j)&= \frac{\sigma_j}{\phi^{1/2}\theta(d_j)}(1 - d_j^{-2}) \text{,  } \theta(d_j) = \phi^{1/2} + \phi^{-1/2} + d_j + d_j^{-1}.  
\end{align*}
% %-------------------------------------------------------------------------------------------------------------------------------------------------------------------------------------
% %-------------------------------------------------------------------------------------------------------------------------------------------------------------------------------------
% %-------------------------------------------------------------------------------------------------------------------------------------------------------------------------------------
The following lemma describes the behaviour of eigenvectors associated with a subset \( A \subseteq \{1, \dots, K\} \), in comparison with the corresponding deterministic positive quadratic form. 
 \begin{lemma}\label{lm: delocalisation bound for outliers}
    Under Assumptions \ref{assp: T N relation} to \ref{assp: assumption 3}, let $A \subseteq \{1, \dots, K\}$ such that $ d_j =d_A $ for all $j \in A$. Then for any deterministic unit vector $\ww \in \mathbb{R}^{T}$, we have 
\begin{align*}
    \langle\ww, P^{(l)}_A \ww\rangle &= \langle\ww, Z_A \ww\rangle + O_{\prec}(n^{-1/2}), l = 1,2. 
\end{align*}
\end{lemma}
Proof: From Theorem 2.16 of \cite{Bloemendal_etal-2016-PTRF}, we have 
 \begin{align}\label{eq: lemma gdboffb 1}
        \langle\ww, P^{(l)}_A \ww\rangle &= \langle\ww, Z_A \ww\rangle + O_{\prec} \Bigg(\frac{1}{p^{1/2}(\phi^{1/2} + d_A) } \sum_{j \in A}\sigma_j w_j^2 \nonumber\\
        &+\left(1 + \frac{\phi^{1/2}d_A^2}{\phi^{1/2} + d_A} \right)\sum_{j=1}^{p} \frac{\sigma_j w_j^2}{p \nu_j(A)^2}\nonumber\\
        &+ \frac{d_A}{\phi^{1/2} + d_A}\left(\sum_{j \in A} \sigma_j w_j^2 \right)^{1/2}\left( \sum_{ j \notin A} \frac{\sigma_j w_j^2}{p \nu_j (A)^2  } \right)^{1/2} \Bigg). 
    \end{align}

    Recall that $\sigma_j = 1+ \phi^{1/2} d_A$ for $j \in A$. Thus, we have
     \begin{align}\label{eq: lemma gdboffb 2}
        \frac{1}{p^{1/2}(\phi^{1/2} + d_A) } \sum_{j \in A}\sigma_j w_j^2 \lesssim  \frac{\phi^{1/2} d_A}{p^{1/2} d_A } \lesssim \frac{1}{n^{1/2}}.
    \end{align}

    Moreover, we can verify that 
    \begin{align*}
        &\nu_j(A)  \asymp d_A \text{ for } j \in A,\\
        &\nu_j(A) \asymp \max\{d_j, d_A\} \text{ for } j \in \{1, \dots, K\} \setminus A \text{ and }\\
        &\nu_j(A) \asymp d_A \text{ for } j \in \{K+1, \dots, p \}.
    \end{align*}

    Therefore, we have 
    \begin{align}\label{eq: lemma gdboffb 3}
        &\left(1 + \frac{\phi^{1/2}d_A^2}{\phi^{1/2} + d_A} \right)\sum_{j=1}^{p} \frac{\sigma_j w_j^2}{p \nu_j(A)^2}\nonumber \\
         =&\left(1 + \frac{\phi^{1/2}d_A^2}{\phi^{1/2} + d_A} \right) \left( \sum_{j = 1}^{K} \frac{\sigma_j w_j^2}{p \nu_j(A)^2}  +\sum_{j=K+1}^{p} \frac{\sigma_j w_j^2}{p \nu_j(A)^2}\right)\nonumber\\
        \lesssim & \frac{\phi d_A }{p(\phi^{1/2} + d_A)} + \frac{\phi^{1/2}}{p(\phi^{1/2} + d_A)}\nonumber\\
        \lesssim &\frac{1}{n} + \frac{1}{p + (np)^{1/2}d_A}.
    \end{align}

    Finally, 
    \begin{align}\label{eq: lemma gdboffb 4}
        &\frac{d_A}{\phi^{1/2} + d_A}\left(\sum_{j \in A} \sigma_j w_j^2 \right)^{1/2}\left( \sum_{ j \notin A} \frac{\sigma_j w_j^2}{p \nu_j (A)^2  } \right)^{1/2} \nonumber\\
        \lesssim& (\phi^{1/2} d_A)^{1/2}\left(\frac{\phi^{1/2}}{pd_A}\right)^{1/2} \nonumber \\
        %=&\left(\frac{\phi}{p}\right)^{1/2}\nonumber \\
       \lesssim& n^{-1/2}. 
    \end{align}
    Therefore, the lemma is proved by \eqref{eq: lemma gdboffb 1}, \eqref{eq: lemma gdboffb 2}, \eqref{eq: lemma gdboffb 3} and \eqref{eq: lemma gdboffb 4}.
%     %------------------------------------------------------------------------------------------------------------------------------------------------------------------------------
%     %------------------------------------------------------------------------------------------------------------------------------------------------------------------------------
%      %------------------------------------------------------------------------------------------------------------------------------------------------------------------------------
    
    The following lemma states some useful results derived from Lemma \ref{lm: delocalisation bound for outliers}: 
 \begin{lemma}\label{lm: inner product for As}
 Under the conditions of Lemma \ref{lm: delocalisation bound for outliers}, we have 
 %\ar{Todo: Update the rate. Also the signal is not strong enough to give $1$ when $d_i$ does not tend to infinity. But we should still be able to get a gap that differentiate $K^*$ and $K^*+1$ as long as a lower bound can be obtained.  } \ar{Can't quite get the same result as \cite{ke_etal-2023-JASA} since $\mu(d_k)$ could be quite small. Get a result first and perhaps argue why stability matters, because the concentration result is somehow tight and that shows the unreliability of the result even if the spikes are "correctly" identified. Perhaps we can still maintain this result for full generality and strenghen the condition as we need. }
     \begin{align}
        \sum_{k \in A}  \langle\vvv, \tilde{v}_k^{(l)}\rangle^2 &=1 + O_{\prec}(n^{-1/2} +\phi^{1/2}d_A^{-1} + d_A^{-2} ) \text{ and } \label{eq: projection of v on tilde v} \\ %
        \sum_{k \in A}  \langle\xixi^{(l)}, \vvv_k\rangle^2 &= 1+ O_{\prec}(n^{-1/2} +\phi^{1/2}d_A^{-1} + d_A^{-2}) \label{eq: projection of xi on v} 
    \end{align}
    for any $\vvv \in \text{Span}\{\vvv_{j}: j \in A \} $ with $\|\vvv\|=1$ and $\xixi^{(l)} \in \text{Span}\{\tilde{\vvv}^{(l)}_{j}: j \in A \}$ with $\|\xixi^{(l)}\|=1$. 
    %Here 
    %$$ \mu(d_j) = \frac{\sigma_j}{\phi^{1/2}(\phi^{1/2} + \phi^{-1/2} + d_j + d_j^{-1})  }(1 - d_j^{-2}).$$
    Moreover, for $j \notin A $, we have
    \begin{align}
      \sum_{k \in A}  \langle\vvv_j, \tilde{v}_k^{(l)}\rangle^2 &=  O_\prec(n^{-1/2}) \text{ and } \nonumber  \\ %\label{eq: ortho of vj on tilde v}
        \sum_{k \in A}  \langle \tilde{\vvv}_j^{(l)}, \vvv_k\rangle^2 &= O_\prec(n^{-1/2}). \nonumber %\label{eq: ortho of tilde vj on v}
        \end{align}
    
    %Equations \eqref{eq: gen projection of v on tilde v} to \eqref{eq: gen ortho of tilde vj on v} hold at the rate $O_{\prec}(N^{-1/2})$ for any $A \subseteq A^{(ub)}$ that satisfy the conditions in Lemma \ref{lm: gen delocalisation bound for outliers far from bulk}. 
 \end{lemma}
 Proof: We only prove \eqref{eq: projection of v on tilde v} and \eqref{eq: projection of xi on v} to save space as the derivations are similar. We first show \eqref{eq: projection of v on tilde v}. Let $\vvv \in \text{Span}\{\vvv_{i}: i \in A \} $ with length $1$. By Lemma \ref{lm: delocalisation bound for outliers}, for $l = 1,2$, we have 
 \begin{align*}
     \langle\vvv, P_A^{(l)}\vvv \rangle = \langle\vvv, Z_A^{(l)}\vvv \rangle + O_{\prec}(n^{-1/2}).
 \end{align*}
% \ar{to update from here}
Note that

 \begin{align}\label{eq: v PA simplification}
      \langle\vvv, P_A^{(l)}\vvv \rangle &= \langle\vvv, \sum_{k \in A} \tilde{v}_k^{(l)}(\tilde{v}_k^{(l)})^{\top}\vvv \rangle  \nonumber\\
      &=\sum_{k \in A}  \langle\vvv, \tilde{v}_k^{(l)}(\tilde{v}_k^{(l)})^{\top}\vvv \rangle \nonumber \\
      & = \sum_{k \in A}  \langle\vvv, \tilde{v}_k^{(l)}\rangle^2.  
 \end{align}
 On the other hand, 
 \begin{align}
     \langle\vvv, Z_A^{(l)}\vvv \rangle &=  \sum_{j \in A} \mu(d_j) \langle\vvv_j,\vvv\rangle^2. \label{eq: v ZA simpliciation} 
 \end{align}
Note that 
\begin{align*}
    \mu(d_j) &= \frac{\sigma_j}{\phi^{1/2}(\phi^{1/2} + \phi^{-1/2} + d_j + d_j^{-1})  }(1 - d_j^{-2})\\
     &= \left(1-\frac{\phi + \phi^{1/2}d_j^{-1}}{\phi + 1 + \phi^{1/2}d_j + \phi^{1/2}d_j^{-1}  }\right)(1 - d_j^{-2})\\
            &= 1 - O_{\prec}(\phi^{1/2}d_j^{-1} + d_j^{-2} ).
\end{align*}
 Hence \eqref{eq: projection of v on tilde v} is proved by \eqref{eq: v PA simplification}, \eqref{eq: v ZA simpliciation} and the fact that $\sum_{j \in A} \langle\vvv_j,\vvv\rangle^2 =1$. 
%  Recall that $ d_j  \geq (1 + \phi^{-1/2}\tau_N )$ for $j \in A^{(b)}$. 
%  Therefore, 
%  \begin{align*}
%     (1 - d_j^{-2}) \geq 1 - \frac{1}{(1 + \phi^{-1/2}\tau_N )^2}\\
%                    = \frac{2\phi^{-1/2}\tau_N + (\phi^{-1/2}\tau_N)^2}{(1 + \phi^{-1/2}\tau_N )^2}
%  \end{align*}
% Since
%  \begin{align*}
%     \frac{\sigma_A}{\phi^{1/2}(\phi^{1/2} + \phi^{-1/2} + d_A + d_A^{-1})  }&= \frac{1 + \phi^{1/2}d_A}{\phi^{1/2}(\phi^{1/2} + \phi^{-1/2} + d_A + d_A^{-1})  }, 
%  \end{align*}
%    it is easy to verify that 
%    \begin{align}\label{eq: mu(d_a) bound}
%        \mu(d_A) = 1 - O_{\prec}(N^{-1}).
%    \end{align}
%    Therefore, \eqref{eq: v PA simplification}, \eqref{eq: v ZA simpliciation} and \eqref{eq: mu(d_a) bound} give 
%     \begin{align*}
%         \sum_{k \in A}  \langle\vvv, \tilde{v}_k^{(l)}\rangle^2 = 1 + O_{\prec}(N^{-1/2} + T^{-1}).
%     \end{align*}
    Following similar argument in Example 2.15 of \cite{Bloemendal_etal-2016-PTRF}, we can interchange the role of $\{\vvv_i\}_{i \in A}$ and $\{\tilde{\vvv}^{(l)}_{i}\}_{i \in A}$ to get \eqref{eq: projection of xi on v}, and the proof is complete.
    % \begin{align}
    %     \sum_{k \in A}  \langle\xixi^{(l)}, \vvv_k\rangle^2 = 1 + O_{\prec}(N^{-1/2} + T^{-1}) \label{eq: projection of xi on v}. %, \\
    %      % \langle\tilde{\vvv}^{(l)}_i, \sum_{k \in A} \vvv_{k}\vvv_{k}^{\top}  \tilde{\vvv}^{(l)}_j \rangle = O_{\prec}(N^{-1/2} + T^{-1}) \text{ for } i \neq j \text{ and }\\
    %      % \langle\vvv_i, \sum_{k \in A} \tilde{\vvv}^{(l)}_{k}(\tilde{\vvv}^{(l)}_{k})^{\top} \vvv_j \rangle = O_{\prec}(N^{-1/2} + T^{-1}) \text{ for } i \neq j. 
    % \end{align}
    % for any $\xixi^{(l)} \in \text{Span}\{\tilde{\vvv}^{(l)}_{i}: i \in A \}$ with $\|\xixi^{(l)}\|=1$.
% %     %--------------------------------------------------------------------------------------------------------------------------------------------------------------------
% %     %--------------------------------------------------------------------------------------------------------------------------------------------------------------------
% %     %--------------------------------------------------------------------------------------------------------------------------------------------------------------------
    
    We proceed to generalise the result of Lemma \ref{lm: inner product for As} to $\{1, \dots, K\}$. Define a partition of $\{1, \dots, K\}$ by $\cup_{s=1}^{S} A_s$, where $S$ is the number of distinct values in $\{\sigma_1, \dots, \sigma_{K}\}$. 
    Specifically, define 
     \begin{align*}
            A_1 &= \{j : \sigma_j = \sigma_1\} \text{ and }\\
            A_s &= \{j : \sigma_j = \sigma_{\sum_{j=1}^{s-1} |A_j|+1 } \} \text{ for } s = 2, \dots, S.
     \end{align*}
For $l = 1,2 $, pick any $\xi^{(l)} \in \text{Span}\{\tilde{\vvv}^{(l)}_1,\dots, \tilde{\vvv}^{(l)}_{K}\}$. We can write
\begin{align*}
    \xi^{(l)} = \sum_{s=1}^{S}\xi_{s}^{(l)},  
\end{align*} 
where $\xi_{s}^{(l)} \in \text{Span}\{\tilde{\vvv}^{(l)}_{j}: j \in A_s \}$. We have the following Lemma:
    \begin{lemma}\label{lm: inner product for whole space}
        Under Assumption \ref{assp: T N relation} to \ref{assp: assumption 3}, for $l = 1, 2$, for any $\xi^{(l)} \in \text{Span}\{\tilde{\vvv}^{(l)}_{1}, \dots, \tilde{\vvv}^{(l)}_{K}\}$, we have
        \begin{align*}
           \max_{\vvv  \in \text{Span}\{\vvv_{1}, \dots, \vvv_{K}\}, \|\vvv\|=1 } \langle\xi^{(l)}, \vvv \rangle = 1  + O_{\prec}(n^{-1/2} +\phi^{1/2}d_K^{-1} + d_K^{-2}).
        \end{align*}
    \end{lemma}
    Proof: From Lemma \ref{lm: inner product for As}, for $j \notin A_s$, we have
\begin{align*}
     \sum_{k \in A_s}\langle\tilde{\vvv}^{(l)}_j, \vvv_k \rangle^2 = O_{\prec}(n^{-1/2}).
\end{align*}
Therefore, we have
\begin{align*}
   \sum_{k \in A_s} \langle\xixi^{(l)},\vvv_k\rangle^2&= \sum_{k \in A_s} \langle\xi_{s}^{(l)},\vvv_k\rangle^2 + O_{\prec}(n^{-1/2}).
\end{align*}
Hence we have 
\begin{align*}
    \sum_{k = 1}^{K} \langle\xixi^{(l)} ,\vvv_k\rangle^2 & = \sum_{s=1}^{S}\sum_{k \in A_s} \langle\xi_{s}^{(l)},\vvv_k\rangle^2 + O_{\prec}(n^{-1/2})\\
                                                &=  \sum_{s=1}^{S} \|\xi_{s}^{(l)}\|^2   + O_{\prec}(n^{-1/2} +\phi^{1/2}d_K^{-1} + d_K^{-2})\\
                                                 &=  1   + O_{\prec}(n^{-1/2} +\phi^{1/2}d_K^{-1} + d_K^{-2})
\end{align*}
by Lemma \ref{lm: inner product for As}. Finally, since $ \langle\xi^{(l)}, \vvv \rangle$ is maximised by taking 
$$\vvv =  \frac{\sum_{k =1}^{K} \langle\xixi^{(l)},\vvv_k\rangle\vvv_k}{(\sum_{k =1}^{K} \langle\xixi^{(l)} ,\vvv_k\rangle^2)^{1/2}} , $$
we have 
\begin{align*}
       \max_{\vvv  \in \text{Span}\{\vvv_{1}, \dots, \vvv_{K^*}\}, \|\vvv\|=1 } \langle\xi^{(l)}, \vvv \rangle  &= \frac{\sum_{k = 1}^{K} \langle\xixi^{(l)} ,\vvv_k\rangle^2}{(\sum_{k =1}^{K} \langle\xixi^{(l)} ,\vvv_k\rangle^2)^{1/2}} \\
       &= 1 +  O_{\prec}(n^{-1/2} +\phi^{1/2}d_K^{-1} + d_K^{-2})
\end{align*}
by taylor's expansion. Thus the proof is complete. \\
We are now ready to prove \eqref{eq: sin angle for k = K} and complete the proof of Proposition \ref{prop: sine angle behaviour}:\\
% % %----------------------------------------------------------------------------------------------------------------------------------------------------------------------------------------
% % %----------------------------------------------------------------------------------------------------------------------------------------------------------------------------------------
% % %----------------------------------------------------------------------------------------------------------------------------------------------------------------------------------------
From Lemma \ref{lm: inner product for whole space}, for $l = 1,2 $, we have
\begin{align*}
    \min_{\xixi^{(l)} \in \tilde{V}^{(l)}_{K} ; \|\xixi^{(l)}\|=1} \max_{\vvv \in V_{K}; \|\vvv\|=1 } \langle\xixi^{(l)},\vvv\rangle 
    =     1 +  O_{\prec}(n^{-1/2} +\phi^{1/2}d_K^{-1} + d_K^{-2}).
\end{align*}
This implies 
\begin{align}\label{eq: equation 1 proof of sin angle}
    \max_{\xixi^{(l)} \in \tilde{V}^{(l)}_{K}; \xixi^{(l)} \neq 0} \min_{\vvv \in V_{K}; \vvv \neq 0} \sin \angle (\xixi^{(l)},\vvv)  =O_{\prec}(n^{-1/2} +\phi^{1/2}d_K^{-1} + d_K^{-2}). 
\end{align}
Note that the roles of $\xixi^{(l)}$ and $\vvv$ in Lemma \ref{lm: inner product for whole space} are interchangeable, as demonstrated in the proof. Hence, by the swapped version of \eqref{eq: equation 1 proof of sin angle}, we can show that
\begin{align*}%\label{eq: equation 2 in lemma 5}
    \max_{\vvv \in V_{K}; \vvv \neq 0} \min_{\xixi^{(l)} \in \tilde{V}^{(l)}_{K}; \xixi^{(l)} \neq 0 } \sin \angle (\vvv,\xixi^{(l)})  = O_{\prec}(n^{-1/2} +\phi^{1/2}d_K^{-1} + d_K^{-2}) \text{ for $l = 1,2$.}
\end{align*}
Therefore, the proof is complete. 
\begin{remark}
   When the eigenvalues are distinct, for $k= 1, \dots, K-1$, it is easy to derive result analogous to Lemma \ref{lm: inner product for whole space} for $\xi^{(l)} \in \text{Span}\{\vvv_1,\dots, \vvv_k\}$. Hence we can verify that \eqref{eq: sin angle for k = K} holds. 
\end{remark}
%--------------------------------------------------------------------------------------------------------------------------------------------------------------------------------------
%--------------------------------------------------------------------------------------------------------------------------------------------------------------------------------------
%--------------------------------------------------------------------------------------------------------------------------------------------------------------------------------------
\subsection{Proof of Theorem \ref{thm: selection for deterministic cn}}
Proof: Without loss of generality, we prove under the assumption that $J=1$. It suffices to show that $$ c_k + \sin \angle (\tilde{V}^{(1)}_{k}, \tilde{V}^{(2)}_{k} ) \gtrsim c_K + \sin \angle (\tilde{V}^{(1)}_{K}, \tilde{V}^{(2)}_{K} ) + \delta/2 \text{ for } k \in \mathcal{K} , k \neq K$$ as $n \to \infty$.

For $k \leq K-1$, since $c_{k} - c_{k+1} >  \delta$, it follows that $c_{k} - c_{K} > (K - k) \delta$. Hence, we have  
\begin{align}\label{eq: proof prop 2 eq 1}
         &c_k + \sin \angle (\tilde{V}^{(1)}_{k}, \tilde{V}^{(2)}_{k} ) - c_{K} - \sin \angle (\tilde{V}^{(1)}_{K}, \tilde{V}^{(2)}_{K} ) \nonumber \\
         >& (K-k)\delta - O_{\prec}(n^{-1/2} +\phi^{1/2}d_K^{-1} + d_K^{-2}) \nonumber \\
         \gtrsim& \delta/2
\end{align}
  as $n \to \infty$ by Proposition \ref{prop: sine angle behaviour}.

  On the other hand, for $k > K$, by Proposition \ref{prop: sine angle behaviour}, we have
  \begin{align}\label{eq: proof prop 2 eq 2}
     & c_k + \sin \angle (\tilde{V}^{(1)}_{k}, \tilde{V}^{(2)}_{k} ) - c_{K} - \sin \angle (\tilde{V}^{(1)}_{K}, \tilde{V}^{(2)}_{K} ) \nonumber \\
     =&c_k -  c_{K} + 1 - O_{\prec}({(p \min\{n, p\})}^{-1/2} ) \nonumber \\
     >& \delta - 1 + 1 - O_{\prec}({(p \min\{n, p\})}^{-1/2} ) \nonumber \\
     \gtrsim& \delta/2
 \end{align}
  as $n \to \infty$. 
Combining \eqref{eq: proof prop 2 eq 1} and \eqref{eq: proof prop 2 eq 2}, the proof is complete. 
%---------------------------------------------------------------------------------------------------------------------------------------------------------------------------------------
%---------------------------------------------------------------------------------------------------------------------------------------------------------------------------------------
%---------------------------------------------------------------------------------------------------------------------------------------------------------------------------------------
\subsection{Proof of Corollary \ref{cor 1}}
Recall that 
\begin{align*}
    \textsc{\textsc{SC2}}(k)  =  \frac{l(k)}{ l(0)}    + \text{INS}(k), 
\end{align*}
where $l(k) = \sum_{j=k+1}^{K_{\max}}\log(\tilde{\sigma}_j +1 )$ for $k = 0, 1, \dots, K_{\max}-1$ and  $l(K_{\max}) =0$. 
 It is obvious that $1 > l(1)/l(0) > \dots >  l(K_{\max})/l(1)=0$.  Moreover, from the assumption in the theorem, we have 
\begin{align}\label{eq: log order}
    \log(\sigma_K)/\log(\sigma_1) \gtrsim 1/C.
\end{align}
From Theorem 2.3 of \cite{Bloemendal_etal-2016-PTRF}, for $j \in \{1, \dots, K\}$, we have %\ar{Self-remark: If $n >>p$, there is the risk that $\tilde{\sigma}_j$ is infinite but $\sigma_j$ is finite, so the bound becomes useless.}
\begin{align*}
    | \tilde{\sigma}_j - \phi^{1/2} - \phi^{-1/2} - d_j - d_j^{-1} | \prec \left(1 + \frac{d_j}{1 + \phi^{-1/2}}\right)\min\{p,n\}^{-1/2}.
\end{align*}
Since $\sigma_j \asymp \phi^{1/2} d_j $, by \eqref{eq: log order} and the assumption that $p \asymp n$, we have %\ar{Should be straightforward}
\begin{align*}
     \log(\tilde{\sigma}_K+1)/\log(\tilde{\sigma}_1+1) \gtrsim 1/C_2
\end{align*}
for some $C_2 >1$. 
This implies that for $k \in \{1, \dots, K-1\}$,
\begin{align*}
\frac{l(k)}{l(0)} - \frac{l(k+1)}{l(0)} & =\frac{\log(\tilde{\sigma}_{k+1} +1)}{l(0)}  \gtrsim \frac{1}{K_{\max}C_2}. 
\end{align*}

On the other hand, 
\begin{align*}
    \frac{l(K)}{l(0)} -\frac{l(K_{\max})}{l(0)}
    \leq& \frac{l(1)}{l(0)} -\frac{l(K_{\max})}{l(0)} 
    = 1 - \frac{ \log(\tilde{\sigma}_1+1)}{ \sum_{j=1}^{K_{\max}}\log(\tilde{\sigma}_j+1 )}
    < 1 - \frac{ 1}{ K_{\max}}.
\end{align*}
Hence the proof is complete by Theorem \ref{thm: selection for deterministic cn}, taking $\delta = K_{\max}C_2$. 
\subsection{Proof of Corollary \ref{cor 2}}
Recall that 
\begin{align*}
    \textsc{SC3}(k) = \frac{\log(1+p^{-1} \sum_{j = k+1}^{p} \tilde{\sigma}_j^2 )}{\log(1+p^{-1} \sum_{j = 1}^{p} \tilde{\sigma}_j^2 )} + \text{INS}(k).
\end{align*}
It is obvious that the first term of $\textsc{SC3}$ lies between $0$ and $1$ for $k \in \mathcal{K}$. Note that $\sigma_k^2 \asymp p$ for $k = 1, \dots, K$. Therefore, by Theorem 2.3 of \cite{Bloemendal_etal-2016-PTRF} and the assumption that $p \asymp n$, we can show that $\tilde{\sigma}_k^2 \asymp p$ for $k = 1, \dots, K$. Let $L(0) = \log(1+p^{-1} \sum_{j = 1}^{p} \tilde{\sigma}_j^2 )$. For $k \in \{1, \dots, K-1\}$, we have
\begin{align}\label{eq: cor 2 lower bound }
\frac{\log(1+ p^{-1} \sum_{j = k+1}^{p} \tilde{\sigma}_j^2  )}{L(0)} -  \frac{\log(1 + p^{-1} \sum_{j = k+2}^{p} \tilde{\sigma}_j^2  )}{L(0)}
     = & \frac{1}{L(0)} \log \left( 1 + \frac{p^{-1}\tilde{\sigma}_{k+1}^2}{1+ p^{-1} \sum_{j = k+2}^{p} \tilde{\sigma}_j^2 }  \right) \nonumber\\
    \geq & \frac{1}{L(0)} \log \left(1 + \frac{p^{-1}\tilde{\sigma}_{K}^2}{1+p^{-1} \sum_{j = 1}^{p} \tilde{\sigma}_j^2 }  \right)>0.
\end{align}
On the other hand,
% \begin{align}\label{eq: cor 2 upper bound }
% \frac{\log(1+p^{-1} \sum_{j = K+1}^{p} \tilde{\sigma}_j^2 )}{L(0)} -  \frac{\log(1+p^{-1} \sum_{j = K_{\max}+1}^{p} \tilde{\sigma}_j^2 )}{L(0)} &\leq 1 - \frac{\log(1+p^{-1} \sum_{j = K_{\max}+1}^{p} \tilde{\sigma}_j^2 )}{L(0)}<1. 
% \end{align}
\begin{align}\label{eq: cor 2 upper bound }
&\frac{\log(1+p^{-1} \sum_{j = K+1}^{p} \tilde{\sigma}_j^2 )}{L(0)} -  \frac{\log(1+p^{-1} \sum_{j = K_{\max}+1}^{p} \tilde{\sigma}_j^2 )}{L(0)} \nonumber \\
\leq& \frac{\log(1+p^{-1} \sum_{j = K+1}^{p} \tilde{\sigma}_j^2 )}{L(0)} \nonumber\\
\leq& 1 - \frac{L(0) - \log(1+p^{-1} \sum_{j = K+1}^{p} \tilde{\sigma}_j^2 )}{L(0)} \nonumber \\
\leq&1 - \frac{1}{L(0)} \log\left(1 + \frac{p^{-1}\tilde{\sigma}_1^2}{1+p^{-1} \sum_{j = K+1}^{p} \tilde{\sigma}_j^2} \right) 
<1.
\end{align}
Hence by \eqref{eq: cor 2 lower bound } and \eqref{eq: cor 2 upper bound }, the conditions of Theorem \ref{thm: selection for deterministic cn} are satisfied and proof is complete. 
% \bibliographystyle{apalike}
% \bibliography{reference}

%\end{document}

\bibliographystyle{apalike}
\bibliography{reference}

\begin{thebibliography}{}

\bibitem[Ahn and Horenstein, 2013]{Ahn_Horenstein-2013-Econometrica}
Ahn, S.~C. and Horenstein, A.~R. (2013).
\newblock Eigenvalue ratio test for the number of factors.
\newblock {\em Econometrica}, 81(3):1203--1227.

\bibitem[Anderson and Rubin, 1956]{anderson1956statistical}
Anderson, T. and Rubin, H. (1956).
\newblock Statistical inference in factor analysis.
\newblock In {\em Proceedings of the Third Berkeley Symposium on Mathematical Statistics and Probability, 1956}, pages 111--150. University of California Press.

\bibitem[Bai, 2003]{bai-2003-Econometrica}
Bai, J. (2003).
\newblock Inferential theory for factor models of large dimensions.
\newblock {\em Econometrica}, 71(1):135--171.

\bibitem[Bai and Ng, 2002]{Bai_Ng-2002-Econometrica}
Bai, J. and Ng, S. (2002).
\newblock Determining the number of factors in approximate factor models.
\newblock {\em Econometrica}, 70(1):191--221.

\bibitem[Bai et~al., 2018]{Bai_etal-2018-Aos}
Bai, Z., Choi, K.~P., and Fujikoshi, Y. (2018).
\newblock Consistency of {AIC} and {BIC} in estimating the number of significant components in high-dimensional principal component analysis.
\newblock {\em The Annals of Statistics}, 46(3):1050--1076.

\bibitem[Bloemendal et~al., 2016]{Bloemendal_etal-2016-PTRF}
Bloemendal, A., Knowles, A., Yau, H.-T., and Yin, J. (2016).
\newblock On the principal components of sample covariance matrices.
\newblock {\em Probability Theory and Related Fields}, 164(1):459--552.

\bibitem[Chen and Li, 2022]{chen_Li-2022-Biometrika}
Chen, Y. and Li, X. (2022).
\newblock Determining the number of factors in high-dimensional generalized latent factor models.
\newblock {\em Biometrika}, 109(3):769--782.

\bibitem[Chen et~al., 2020]{Chen_Li_Zhang-2020-JASA}
Chen, Y., Li, X., and Zhang, S. (2020).
\newblock Structured latent factor analysis for large-scale data: Identifiability, estimability, and their implications.
\newblock {\em Journal of the American Statistical Association}, 115(532):1756--1770.

\bibitem[Danziger et~al., 2006]{Danziger_etal-2006-IEEE}
Danziger, S.~A., Swamidass, S.~J., Zeng, J., Dearth, L.~R., Lu, Q., Chen, J.~H., Cheng, J., Hoang, V.~P., Saigo, H., Luo, R., et~al. (2006).
\newblock Functional census of mutation sequence spaces: the example of p53 cancer rescue mutants.
\newblock {\em IEEE/ACM Transactions on Computational Biology and Bioinformatics}, 3(2):114--125.

\bibitem[Dobriban and Owen, 2019]{dobriban_owen-2019-JRSSB}
Dobriban, E. and Owen, A.~B. (2019).
\newblock Deterministic parallel analysis: an improved method for selecting factors and principal components.
\newblock {\em Journal of the Royal Statistical Society Series B: Statistical Methodology}, 81(1):163--183.

\bibitem[Fang and Wang, 2012]{fang2012selection}
Fang, Y. and Wang, J. (2012).
\newblock Selection of the number of clusters via the bootstrap method.
\newblock {\em Computational Statistics \& Data Analysis}, 56(3):468--477.

\bibitem[Ke et~al., 2023]{ke_etal-2023-JASA}
Ke, Z.~T., Ma, Y., and Lin, X. (2023).
\newblock Estimation of the number of spiked eigenvalues in a covariance matrix by bulk eigenvalue matching analysis.
\newblock {\em Journal of the American Statistical Association}, 118(541):374--392.

\bibitem[Lim and Yu, 2016]{lim2016estimation}
Lim, C. and Yu, B. (2016).
\newblock Estimation stability with cross-validation ({ESCV}).
\newblock {\em Journal of Computational and Graphical Statistics}, 25(2):464--492.

\bibitem[Liu et~al., 2010]{liu2010stability}
Liu, H., Roeder, K., and Wasserman, L. (2010).
\newblock Stability approach to regularization selection (stars) for high dimensional graphical models.
\newblock {\em Advances in Neural Information Processing Systems}, 23.

\bibitem[Liu et~al., 2023]{Liu_etal-2023-Psychometrika}
Liu, X., Wallin, G., Chen, Y., and Moustaki, I. (2023).
\newblock Rotation to sparse loadings using {$L^P$} losses and related inference problems.
\newblock {\em Psychometrika}, 88:527--553.

\bibitem[Lopes et~al., 2019]{Lopes_etal-2019-Biometrika}
Lopes, M.~E., Blandino, A., and Aue, A. (2019).
\newblock Bootstrapping spectral statistics in high dimensions.
\newblock {\em Biometrika}, 106(4):781--801.

\bibitem[Onatski, 2009]{Onatski-2009-Econometrica}
Onatski, A. (2009).
\newblock Testing hypotheses about the number of factors in large factor models.
\newblock {\em Econometrica}, 77(5):1447--1479.

\bibitem[O'Rourke et~al., 2018]{ORourke_etal-2018-LAA}
O'Rourke, S., Vu, V., and Wang, K. (2018).
\newblock Random perturbation of low rank matrices: Improving classical bounds.
\newblock {\em Linear Algebra and its Applications}, 540:26--59.

\bibitem[Pfister et~al., 2021]{pfister2021stabilizing}
Pfister, N., Williams, E.~G., Peters, J., Aebersold, R., and B{\"u}hlmann, P. (2021).
\newblock Stabilizing variable selection and regression.
\newblock {\em The Annals of Applied Statistics}, 15(3):1220--1246.

\bibitem[Rohe and Zeng, 2023]{Rohe_Zeng-2023-JRSSB}
Rohe, K. and Zeng, M. (2023).
\newblock Vintage factor analysis with varimax performs statistical inference.
\newblock {\em Journal of the Royal Statistical Society Series B: Statistical Methodology}, 85:1037--1060.

\bibitem[Sun et~al., 2013]{sun2013consistent}
Sun, W., Wang, J., and Fang, Y. (2013).
\newblock Consistent selection of tuning parameters via variable selection stability.
\newblock {\em The Journal of Machine Learning Research}, 14(1):3419--3440.

\bibitem[Sun et~al., 2016]{sun2016stabilized}
Sun, W.~W., Qiao, X., and Cheng, G. (2016).
\newblock Stabilized nearest neighbor classifier and its statistical properties.
\newblock {\em Journal of the American Statistical Association}, 111(515):1254--1265.

\bibitem[Wang, 2010]{wang2010consistent}
Wang, J. (2010).
\newblock Consistent selection of the number of clusters via crossvalidation.
\newblock {\em Biometrika}, 97(4):893--904.

\bibitem[Yu, 2013]{yu2013stability}
Yu, B. (2013).
\newblock Stability.
\newblock {\em Bernoulli}, 19(4):1484--1500.

\bibitem[Yu and Kumbier, 2020]{yu2020inaugural}
Yu, B. and Kumbier, K. (2020).
\newblock Inaugural article by a recently elected academy member: Veridical data science.
\newblock {\em Proceedings of the National Academy of Sciences of the United States of America}, 117(8):3920.

\bibitem[Yu et~al., 2015]{yu2015useful}
Yu, Y., Wang, T., and Samworth, R.~J. (2015).
\newblock A useful variant of the {Davis--Kahan} theorem for statisticians.
\newblock {\em Biometrika}, 102(2):315--323.

\end{thebibliography}

\end{document}